\documentclass[aps,prl,floatfix,superscriptaddress,nofootinbib,
amsmath,amssymb,twocolumn,preprintnumbers]{revtex4-2}
\pdfoutput=1
\usepackage[dvipsnames]{xcolor}
\usepackage{graphicx}
\usepackage[colorlinks=true,citecolor=blue,linkcolor=blue,breaklinks=true]{hyperref}
\usepackage{soul}
\usepackage{mathtools}
\usepackage{multirow}
\usepackage{gensymb}
\usepackage{orcidlink}

\def\bea{\begin{eqnarray}}
\def\eea{\end{eqnarray}}
\def\beq{\begin{equation}}
\def\eeq{\end{equation}}

\begin{document}

\title{Probing Feebly Interacting Particles with 511 keV Line from Circumstellar Medium of Supernovae 
}

\author{Yu Cheng
\orcidlink{0000-0002-4822-3890}}
\email{chengyu059@uchicago.edu}
\affiliation{Department of Physics, Enrico Fermi Institute, Leinweber Institute for Theoretical Physics, Kavli Institute for Cosmological Physics, University of Chicago, Chicago, IL 60637, USA}
\affiliation{Department of Physics, Korea Advanced Institute of Science and Technology (KAIST), Daejeon 34141, Korea}

\author{Chui-Fan Kong
\orcidlink{0009-0007-7010-5085}}
\email{kongcf@ibs.re.kr}
\affiliation{Particle Theory and Cosmology Group (PTC), Center for Theoretical Physics of the Universe (CTPU),
Institute for Basic Science, Daejeon 34126, Korea}

\author{Yen-Hsun Lin
\orcidlink{0000-0001-7911-7591}}
\email{yenhsun@phys.ncku.edu.tw}
\affiliation{Institute of Physics, Academia Sinica, Taipei 115, Taiwan}

\author{Meng-Ru Wu
\orcidlink{0000-0003-4960-8706}}
\email{mwu@as.edu.tw}
\affiliation{Institute of Physics, Academia Sinica, Taipei 115, Taiwan}
\affiliation{Institute of Astronomy and Astrophysics, Academia Sinica, Taipei 106, Taiwan}
\affiliation{Physics Division, National Center for Theoretical Sciences, Taipei 106, Taiwan}

\author{Seokhoon Yun
\orcidlink{0000-0002-7960-3933}}
\email{seokhoon.yun@knu.ac.kr}
\affiliation{Department of Physics, Kyungpook National University, Daegu 41566, Korea}
\affiliation{Particle Theory and Cosmology Group (PTC), Center for Theoretical Physics of the Universe (CTPU),
Institute for Basic Science, Daejeon 34126, Korea}

\begin{abstract}
Core-collapse supernovae (CCSNe) are intense sources of feebly interacting particles (FIPs), whose visible decays beyond the stellar envelope can generate distinct electromagnetic signals.
We propose the circumstellar medium (CSM) surrounding the CCSNe progenitor as a new target for 511 keV line searches. 
FIPs escaping the stellar interior may decay into electron–positron pairs and inject positrons into the CSM, where they slow down and subsequently annihilate, giving rise to a 511 keV line as an electromagnetic precursor to the supernova shock breakout.
Using dark photons as a benchmark, we calculate the time-dependent signal for CSM profiles motivated by SN 2023ixf and SN 2024ggi. 
For a Galactic supernova at 10 kpc, the projected reach of COSI and AMEGO extends beyond existing supernova limits over part of the MeV-scale dark-photon parameter space.
The same FIP decays can simultaneously heat the CSM and sublimate circumstellar dust, producing a rapid spectral transition from infrared excess to optical or ultraviolet emission~\cite{Cheng:2026ugs}.
The coincident appearance of this transition and the 511~keV line would provide a distinctive correlated multiwavelength signature of FIP decays in the dense presupernova environment.
\end{abstract}

\preprint{CTPU-PTC-26-26}

\maketitle

\textit{\textbf{Introduction.}---}
A proto-neutron star (PNS) born during a core-collapse supernova (CCSN) explosion reaches temperatures of $\mathcal{O}(30)\,{\rm MeV}$ and densities exceeding nuclear saturation, making it an efficient source of feebly interacting particles (FIPs), including axions, sterile neutrinos, dark photons, and other light dark-sector particles~\cite{Kazanas:2014mca,Arguelles:2016uwb,Jaeckel:2017tud,DeRocco:2019njg,Sung:2019xie,Lucente:2020whw,Caputo:2021rux,Calore:2021lih,Caputo:2022mah,Hoof:2022xbe,Ferreira:2022xlw,Lella:2022uwi,Diamond:2023scc,Chauhan:2023sci,Carenza:2023old,Muller:2023pip,DelaTorreLuque:2024zsr,Mori:2024vrf,Lella:2024dmx,Benabou:2024jlj,Takata:2025lyu,Fiorillo:2025yzf,Chauhan:2025mnn,Balaji:2025alr,Fiorillo:2025sln,Candon:2025ypl,Caputo:2025avc,Ferreira:2025qui,Huang:2025xvo,Blinov:2025aha,Joseph:2026nut,Yu:2026xtb,Raffelt:1987yt,Turner:1987by,Mayle:1987as,Raffelt:1990yz,Janka:1995ir,Raffelt:1996wa,Farzan:2002wx,Raffelt:2011nc,Dent:2012mx,Rrapaj:2015wgs,Chang:2016ntp,Hardy:2016kme,Fischer:2016cyd,Chang:2018rso,Carenza:2019pxu,Bar:2019ifz,Mastrototaro:2019vug,Suliga:2019bsq,Dev:2020eam,Carenza:2020cis,Suliga:2020vpz,Fischer:2021jfm,Sung:2021swd,Shin:2022ulh,Ho:2022oaw,Ray:2023gtu,Lella:2023bfb,Akita:2023iwq,Ray:2024jeu,Manzari:2024jns,Fiorillo:2024upk,Syvolap:2024hdh,Li:2024pcp,Springmann:2024ret,Vogl:2024ack,Alonso-Gonzalez:2024ems,Hardy:2024gwy,Caputo:2025aac,Cappiello:2025tws,Fiorillo:2025gnd,Candon:2025sdm,Mori:2025cqf,Mori:2025eit,Gupta:2025ygk,Chauhan:2025xqr,Caputo:2026pdw,Qiu:2026irb,Cheng:2026ugs}.
The classic CCSN constraint arises from SN~1987A, 
where excessive energy loss of the PNS through FIP emission would have shortened or otherwise distorted the observed neutrino burst~\cite{Raffelt:1996wa,Chang:2016ntp,Rrapaj:2015wgs}.
If these particles subsequently decay into visible final states, their decay products can provide complementary signatures whose form depends on the decay location.
Examples include 
excessive supernova explosion energy, 
prompt $\gamma$-ray signals, diffuse $\gamma$-ray backgrounds, fireball formation, and Galactic positron injection that later contributes to the Galactic 511~keV annihilation line~\cite{Kazanas:2014mca,Arguelles:2016uwb,Jaeckel:2017tud,DeRocco:2019njg,Sung:2019xie,Lucente:2020whw,Caputo:2021rux,Calore:2021lih,Caputo:2022mah,Hoof:2022xbe,Ferreira:2022xlw,Lella:2022uwi,Diamond:2023scc,Chauhan:2023sci,Carenza:2023old,Muller:2023pip,DelaTorreLuque:2024zsr,Mori:2024vrf,Lella:2024dmx,Benabou:2024jlj,Takata:2025lyu,Fiorillo:2025yzf,Chauhan:2025mnn,Balaji:2025alr,Fiorillo:2025sln,Candon:2025ypl,Caputo:2025avc,Ferreira:2025qui,Huang:2025xvo,Blinov:2025aha,Joseph:2026nut,Yu:2026xtb}.

The 511~keV line is a particular target for MeV $\gamma$-ray astronomy. 
COSI is designed to survey the 511~keV sky with high-resolution spectroscopy, while proposed missions such as AMEGO, GECCO, GRAMS, MASS, and newASTROGAM 
would further improve sensitivity across the MeV gap~\cite{Tomsick:2019wvo,Caputo:2017sjw,e-ASTROGAM:2018jlu,Kierans:2020otl,Caputo:2022xpx,Tomsick:2023aue,Orlando:2021get,Aramaki:2019bpi,Zhu:2023zwr,Berge:2025kff}.
Ref.~\cite{Chauhan:2025xqr} investigated the 511~keV line from FIP decays that efficiently produce positrons near the
stellar surface.
Building on the recent proposal 
that considers 
the common existence of the circumstellar medium (CSM) surrounding supernova progenitors as a probe of visible energy deposition from FIP decays~\cite{Cheng:2026ugs}, we present in this \emph{Letter} the first study of a transient 511~keV annihilation signal generated from FIP decays in the CSM. 

\begin{figure}[t!]
\begin{centering}
\includegraphics[width=1.0\columnwidth]{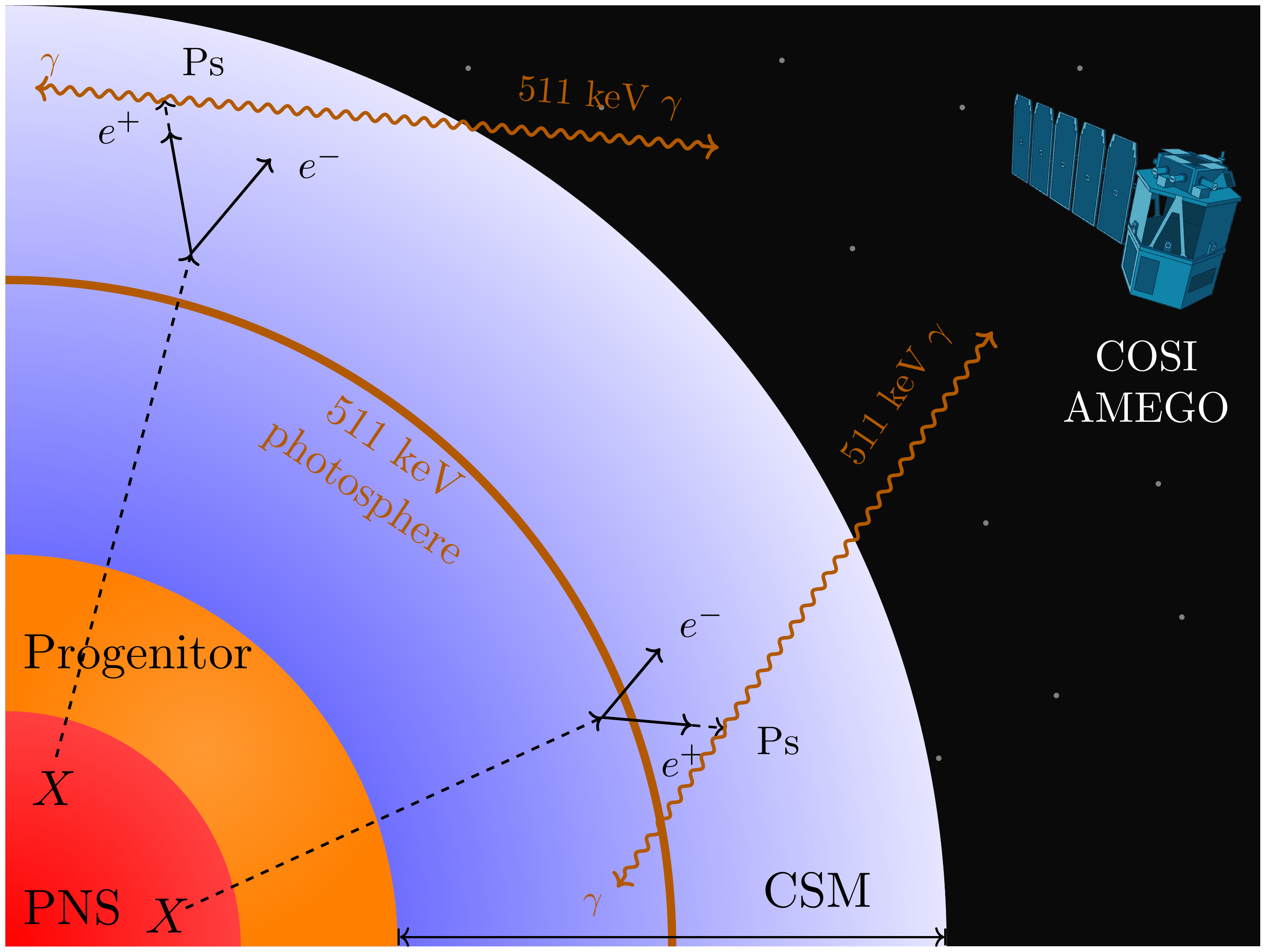}
\end{centering}
\caption{\label{fig:scheme}
Schematic illustration of the proposed 511~keV line signal.
FIPs ($X$) are produced in the PNS and decay into
electron-positron pairs in the CSM.
After losing kinetic energy and stopping in
the CSM, the positrons form positronium and generate 511~keV photons.
Photons produced outside the effective 511~keV photosphere can escape and be observed by $\gamma$-ray telescopes such as COSI and AMEGO.
}
\end{figure}

In what follows, we will show that
positrons injected by 
FIP decays in CSM can efficiently lose their kinetic energy, leading to annihilation with ambient electrons at different stages and producing 511~keV photons.  
Unlike the compact stellar envelope considered in Ref.~\cite{Chauhan:2025xqr}, where the large optical depth restricts the free streaming of 511\,keV photons to a small region, 
a large fraction of produced 511\,keV photons can escape from the relatively diffuse CSM.
These escaping 511~keV photons result in an extended signal lasting for $10^3\text{-}10^5 \mathrm{~s}$ due to geometric time delays across the extended CSM, enabling ongoing and future MeV $\gamma$-ray telescopes to probe uncharted parameter space for FIP when the next galactic CCSN occurs. 
A schematic plot for this process is illustrated in Fig.~\ref{fig:scheme}.
Hereafter we adopt the natural unit, $c=\hbar=1$.

\textit{\textbf{CSM and Benchmark Models.---}}
Stellar mass loss prior to core collapse can 
create extended CSM, whose interaction with the supernova shock and ejecta shapes early supernova light curves and spectra, allowing the CSM density structure to be inferred~\cite{Smith:2014txa,Chandra:2017aev,Smith_2017,2020RSOS....700467F,Dessart:2024mop}.
Recent flash spectroscopy surveys find dense material near the progenitors of more than 30\% of Type-II supernovae, indicating enhanced mass loss shortly before explosion~\cite{Forster:2018mib,Bruch:2020jcr,Jacobson-Galan:2021pki,Bruch:2022aqd}; the resulting CSM can be confined, stratified, or asymmetric~\cite{Vasylyev:2025nsj}. 

We use observationally motivated CSM density profiles for the nearby events SN 2023ixf~\cite{Zimmerman:2023mls,Jacobson-Galan:2023ohh,Jacobson-Galan:2025rbd} and SN 2024ggi~\cite{Jacobson-Galan:2024vbq,Ertini:2025pse,Borowska-Naguszewska:2025que} as our main benchmarks.
They represent a dense inner region up to $r\sim 10^{14-15}$\,cm, corresponding to a mass-loss rate of order $\dot{M} \gtrsim 10^{-2}M_\odot\,{\rm yr}^{-1}$, followed by a lower-density outer component extending to larger radii with $\dot{M} \lesssim 10^{-3}M_\odot{\rm yr}^{-1}$; see  Fig.~\ref{fig:density} in Supplemental Material for their density profiles. 
Additional reference configurations and associated parameters are presented only in Supplemental Material.

For concreteness, we consider 
the dark photon (DP) as an illustrative FIP benchmark throughout 
this work.
Its interactions with the Standard Model are controlled by the dimensionless kinetic-mixing $\varepsilon$~\cite{Okun:1982xi,Galison:1983pa,Holdom:1985ag}, through $\mathcal L \supset (\varepsilon/2) F_{\mu\nu}F^{\prime\mu\nu}$,
where $F_{\mu\nu}$ and $F'_{\mu\nu}$ denote the photon and DP field strengths, respectively.
We focus on the DP mass $m_{\gamma^\prime} > 2 m_e$, for which $\gamma^\prime \to e^+e^-$ is kinematically allowed.
The relevant microphysics, including DP production in the PNS, in-medium mixing effects, and DP visible decay, has been spelled out in the literature; see, e.g., Ref.~\cite{Caputo:2025avc} and references therein. 
We follow Ref.~\cite{Caputo:2025avc} and 
employ the PNS thermodynamic profiles from 
the spherically symmetric SFHo-18.8 CCSN 
model~\cite{Bollig:2020phc,Garching}
to compute the DP production as in Ref.~\cite{Cheng:2026ugs}.

 \textit{\textbf{Formation of 511\,keV Line from DP Decay.}---}
For a DP produced in the PNS with momentum $k$ and energy $\omega=\sqrt{k^2+m_{\gamma^\prime}^2}$, its decay rate into an electron-positron pair in the lab frame reads
\begin{equation}
 \Gamma_{\rm lab}(k)=
 \frac{\alpha \varepsilon^2}{3}
 \frac{m_{\gamma^\prime}^2}{\omega}
 \sqrt{1-\frac{4 m_e^2}{m_{\gamma^\prime}^2}}
 \left(1+\frac{2 m_e^2}{m_{\gamma^\prime}^2}\right)\,.
 \label{eq:gammalab}
\end{equation}

We denote by $r_{\rm d}$ and $t_{\rm d}$ the decay radius and decay time, respectively, measured from core collapse.
Since the PNS size is negligible compared with
the CSM radii of interest, DPs emitted from the PNS can be treated as originating from the center and propagating radially outward.
A DP with momentum $k$ travels with velocity $\beta_{\gamma^\prime} = k/\omega$ and obeys the time-of-flight relation $r_{\rm d} = \beta_{\gamma^\prime} t_{\rm d}$.
The probability density for it to decay at time $t_{\rm d}$ is $dP_{\gamma^\prime}^{(k)}/dt_{\rm d}=\Gamma_{\rm lab}(k)e^{-\Gamma_{\rm lab}(k)t_{\rm d}}$.
The resulting number of DP decays per unit radius and time is
\begin{equation}
\begin{split}
 \frac{d^2 N^{\rm dec}_{\gamma^\prime}}
 { d r_{\rm d}\, d t_{\rm d}}
 &=
\int dk 
    \frac{dN_{\gamma^\prime}}{dk}
    \frac{dP_{\gamma^\prime}^{(k)}}{dt_{\rm d}}
    \delta(r_{\rm d}- \beta_{\gamma'}t_{\rm d})\\
 &=
 \left.\frac{d N_{\gamma^\prime}}{d k}\right|_{k=k_\ast}\,
 \Gamma_{\rm lab}(k_\ast)
 e^{-\Gamma_{\rm lab}(k_\ast)t_{\rm d}}
 \frac{\omega^3_\ast}{t_{\rm d} m_{\gamma^\prime}^2}\,,
\end{split}
\label{eq:decaydistribution}
\end{equation}
where $k_* = m_{\gamma'}r_{\rm d}/
\sqrt{t_{\rm d}^2-r_{\rm d}^2}$ with $0<r_{\rm d}<t_{\rm d}$.
The spectrum $dN_{\gamma^\prime}/dk$ is integrated over the production region near the PNS and over the DP-emission duration, which we take to be $10\,{\rm s}$.
Since this duration is much shorter than the characteristic timescale of the 511~keV line signal,  
we treat DP production as instantaneous at core collapse.
The last term accounts for the Jacobian factor $\omega^3/(t_{\rm d} m_{\gamma^\prime}^2)$ arising from integrating the delta function over $k$.

Assuming $\gamma^\prime\to e^+e^-$ is the dominant DP decay channel, each DP decay injects one $e^+e^-$ pair.
For a DP with momentum $k$, the energy distribution of the injected positron is a flat box spectrum in the lab frame,
\begin{equation}
\begin{aligned}
 \frac{d N_{e^+}}{dE_i}
 &=
 \frac{1}
 {2 \gamma \beta_{\gamma^\prime} p_e}
 \Theta(E_i - E_{-})\Theta(E_{+}-E_i),
\end{aligned}
\label{eq:positron-box}
\end{equation}
where $E_{\pm} = \gamma( m_{\gamma^\prime}/2 \pm \beta_{\gamma^\prime} p_e)$, $\gamma = \omega/m_{\gamma^\prime}$ is the Lorentz factor of the DP, $p_e=\sqrt{m_{\gamma^\prime}^2/4-m_e^2}$ is the positron momentum in the DP rest frame, and $\Theta$ is the Heaviside step function.
Combining Eqs.~\eqref{eq:decaydistribution} and \eqref{eq:positron-box}
gives the differential positron injection rate
\begin{equation}
 Q_+(r_{\rm d},t_{\rm d},E_i)
 \equiv
 \frac{d^3N_{e^+}}{dr_{\rm d}\,dt_{\rm d}\,dE_i}
 =
 \frac{d^2N^{\rm dec}_{\gamma^\prime}}{dr_{\rm d}\,dt_{\rm d}}
 \frac{dN_{e^+}}{dE_i}.
\label{eq:qplus}
\end{equation}

\begin{figure}[t!]
\begin{centering}
\includegraphics[width=1.0\columnwidth]{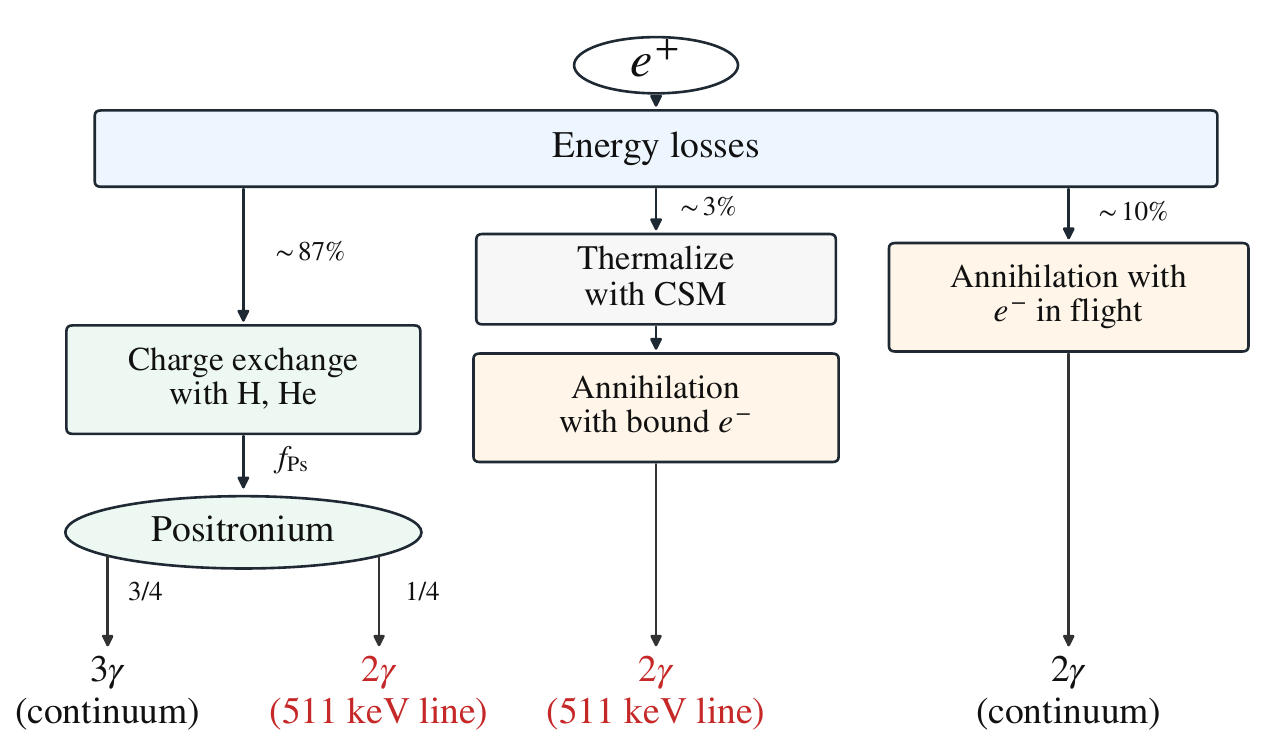}
\end{centering}
\caption{
Fate of positrons injected into a cold and neutral CSM. 
Most surviving positrons
form positronium through charge exchange with H or He~\cite{Guessoum:2005cb,Prantzos:2010wi}. 
The smaller thermalized component can annihilate directly with bound
electrons~\cite{Guessoum:2005cb,Prantzos:2010wi}, 
while annihilation in flight generates a broad continuum photon spectrum~\cite{Beacom:2005qv,Prantzos:2010wi,DelaTorreLuque:2024zsr}. 
The percentages show in the figure indicate branch ratio for each process in the neutral gas benchmark.
If energy deposition substantially heats or ionizes the CSM, these branching fractions can change; see text for discussion.
}
\label{fig:AnnihilationChannels}
\end{figure}

After injection into the CSM, which is assumed to possess a solar-like composition, each positron loses kinetic energy via interactions with the surrounding gas.
Annihilation can occur at different stages (Fig.~\ref{fig:AnnihilationChannels}): a positron may annihilate in flight while slowing down and produce a broad continuum photon spectrum~\cite{Beacom:2005qv,DelaTorreLuque:2024zsr}, or it may survive to low energies and subsequently form positronium or annihilate directly~\cite{Guessoum:2005cb,Prantzos:2010wi}.
We describe the slowing-down stage using the continuous-slowing-down approximation (CSDA) and define $r_{\rm s}(r_{\rm d},E_i)$ as the radius at which a positron injected at $r_{\rm d}$ with energy $E_i$ reaches $E_{\rm stop}=m_e+K_{\rm stop}$.
We take $K_{\rm stop}=10\,{\rm keV}$, below which the low-energy annihilation channels are treated explicitly using Monte-Carlo simulations detailed in the End Matter.
The corresponding stopping distance is $\delta r_{\rm stop}\equiv r_{\rm s}-r_{\rm d}$, and the stopping time is $t_{\rm stop}\simeq\delta r_{\rm stop}$.
Because the positron is nonrelativistic below $K_{\rm stop}$, its residual propagation distance before positronium formation or direct annihilation is negligible compared with the CSDA stopping distance, and we identify $r_{\rm s}$ as the effective location of these processes.

We define $P_{\rm surv}(E_i)$ as the probability that a positron with initial energy $E_i$ reaches $E_{\rm stop}$ without undergoing in-flight annihilation.
For $E_i\in[1,200]\,{\rm MeV}$, we find $P_{\rm surv}=[0.99,0.79]$, implying that only an $\mathcal{O}(10)\%$ fraction of the injected positrons annihilate in flight before reaching $E_{\rm stop}$. 
Below $E_{\rm stop}$, in-flight annihilation is negligible.
For cold CSM ($T\lesssim 3000$~K), most surviving positrons ($\sim 87\%$ 
out of all injected positrons) form positronium through charge exchange with neutral atoms and radiative recombination with free electrons.
The remaining positrons thermalize and annihilate directly with free or bound electrons~\cite{Guessoum:2005cb,Prantzos:2010wi}.
Of the positronium states, the para-positronium component, which constitutes $1/4$ of the total, decays into two 511\,keV photons, whereas ortho-positronium produces a continuum below 511\,keV; see End Matter for details.
Fig.~\ref{fig:AnnihilationChannels} summarizes these branching ratios.
Note that these fractions assume an initially cold CSM; possible modifications due to CSM heating are discussed below.

\textit{\textbf{511\,keV $\gamma$-ray Escape and Observed Flux.---}}
Assuming that all the slow positrons surviving from in-flight annihilation go through charge-exchanges to produce 511\,keV photons, we combine the positron-injection distribution in Eq.~\eqref{eq:qplus} with the in-flight survival probability $P_{\rm surv}(E_i)$, and map each positron to its stopping radius and time to obtain the 511~keV photon production rate per unit radius,
\begin{equation}
\begin{split}
 Q_{511}(r,t)
 &\simeq 
 2 f_{\rm pPs}
 \int dr_{\rm d}\,dt_{\rm d}\,dE_i\,
 Q_+(r_{\rm d},t_{\rm d},E_i)\,
 P_{\rm surv}(E_i)
 \\
 &\quad\times
 \delta\!\left[r-r_{\rm s}(r_{\rm d},E_i)\right]
 \delta\!\left[t-t_{\rm d}-t_{\rm stop}(r_{\rm d},E_i)\right]\,,
\end{split}
\label{eq:q511emissivity}
\end{equation}
where $r$ and $t$ denote the emission radius and time of the 511\,keV photons, 
respectively, $f_{\rm pPs}=1/4$ is the para-positronium fraction, 
and the factor of two accounts for the two photons produced in each para-positronium decay. 
The two delta functions fix the photon-production radius to the positron stopping radius, $r_{\rm s}(r_{\rm d},E_i)$,
and the production time to the sum of the DP decay and positron stopping times.

The produced photons contribute to the observable narrow line only if they escape from the CSM without being absorbed or scattered to other energy range.
For a given CSM density profile $\rho(r)$, the outward optical depth for a 511\,keV photon produced at radius $r$ is $\tau_{511}(r)\equiv \int_{r}^{R_{\rm max}}\kappa_{511}\rho(r')\,dr'$, where $R_{\rm max}$ is the outer boundary of the CSM and $\kappa_{511}$ is the mass attenuation coefficient evaluated at $E_\gamma=511\,{\rm keV}$, which we take from the XCOM tables~\cite{XCOM}.
We define the 511~keV photospheric radius $r^{511}_{\rm ph}$ by taking $\tau_{511}(r^{511}_{\rm ph})=2/3$.

Because of the finite spatial extent of the CSM, 511~keV photons emitted at different positions do not arrive at the observer simultaneously.
Let us consider a positron originating from a DP decay at radius $r_{\rm d}$ and subsequently annihilating to emit a 511~keV photon at the positron stopping radius $r_{\rm s} =  r_{\rm d} + \delta r_{\rm stop}$.
For an emission point at polar angle $\theta$ relative to the line of sight, the observed time is, in the long-distance limit of $D\gg r_{\rm s}$ with $D$ being the distance to SN,
\begin{equation}
t_{\rm obs}
\simeq
\frac{r_{\rm d}}{\beta_{\gamma'}}
+t_{\rm stop}(r_{\rm d},E_i)
-r_{\rm s}\cos\theta \,.
\end{equation}
We set $t_{\rm obs}=0$ as the arrival time of a reference neutrino from the center at core collapse, thereby subtracting the common source-to-observer propagation time $D$.

Assuming spherical symmetry, we retain only emission from the near-side hemisphere, $\cos\theta\in[0,1]$.
Far-side photons can in principle reach the observer, but traverse a larger CSM column and may be obscured by the stellar envelope or ejecta.
Neglecting this contribution therefore yields a conservative estimate.
Then, the observed 511\,keV flux reads
\begin{equation}
\begin{split}
 \Phi_{511}(t_{\rm obs})
 &=
 \frac{1}{8\pi D^2}
 \int_{r_{\rm ph}^{511}}^{R_{\rm max}}dr
 \int_0^1 d \cos\theta
 \\
 &\quad\times
 Q_{511}\!\left(r,t_{\rm obs}  + r \cos\theta\right)\,.
\end{split}
 \label{eq:flux}
\end{equation}
The factor $1/(4\pi D^2)$ accounts for geometric dilution, while the additional factor $1/2$ accounts for the fraction of isotropically emitted photons propagating toward the observer.
Clearly, photons emitted simultaneously from a shell at radius $r$ arrive over $0\leq\Delta t\leq r$  because of path-length differences across the near-side hemisphere.
The observed signal duration is therefore determined by the distributions of DP decay and positron stopping times, together with geometric light-travel-time delays.
For the CSM profiles and DP parameter space considered below, the resulting 511~keV line  
signal typically persists for $\sim10^3$-$10^5\,{\rm s}$.

\begin{figure}[t!]
\begin{centering}
\includegraphics[width=1.0\columnwidth]{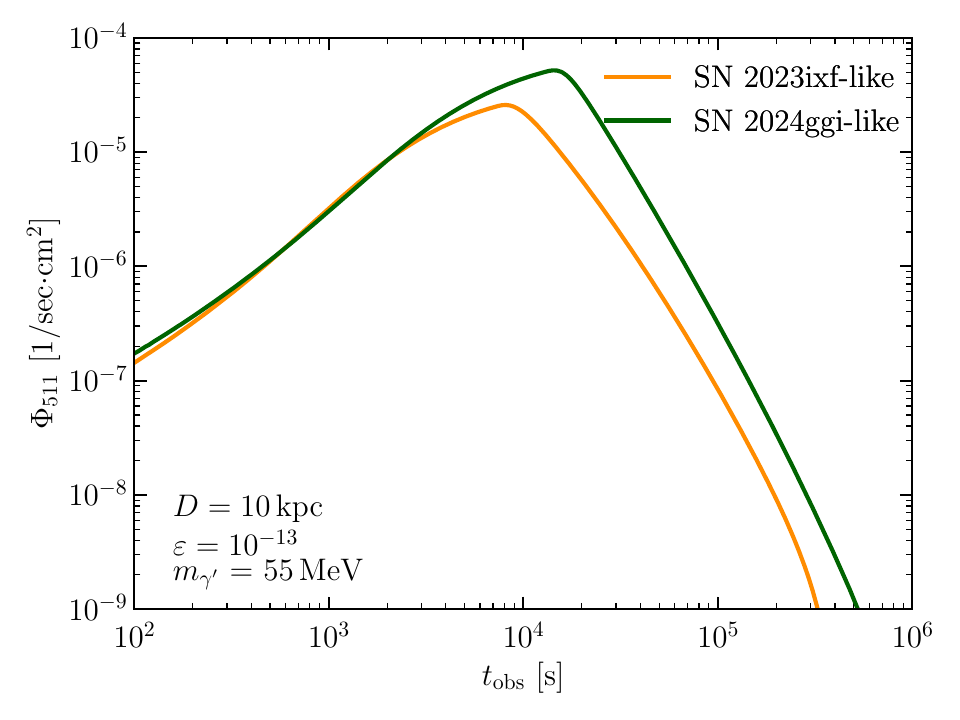}
\end{centering}
\caption{\label{fig:time_spectra}
Time-dependent 511 keV flux for a Galactic CCSN at 
$D=10 \, {\rm kpc}$, with $m_{\gamma^{\prime}}=55 \,{\rm MeV}$ and $\varepsilon=10^{-13}$. 
The solid lines correspond to the two observationally motivated CSM profiles, SN 2023ixf and SN 2024ggi. 
The peak tracks the 511 keV photospheric radius, while the extended tail arises from positrons stopping at larger radii and from geometric light travel time delays.
}
\end{figure}

Fig.~\ref{fig:time_spectra} shows the time-dependent 511 keV $\gamma$-ray flux, $\Phi_{511}(t_{\rm obs})$, for $m_{\gamma^\prime} = 55\,{\rm MeV}$ and $\varepsilon=10^{-13}$ from a Galactic CCSN at a distance $D=10\,{\rm kpc}$ with the two benchmark CSM profiles motivated by SN 2023ixf~\cite{Zimmerman:2023mls} and SN 2024ggi~\cite{Jacobson-Galan:2024vbq}.
The signal peaks at $\sim 10^4\,$s for both profiles with a subdominant tail extending to later times. 
The peak time is set approximately by the light-crossing time of the 511 keV photosphere, $r_{\rm ph}^{511}$,
since a larger fraction of the escaping 511 keV line photons originates from positrons that stop near this radius. 
For SN~2023ixf and SN~2024ggi, $r_{\rm ph}^{511} = 2.18 \times 10^{14}\,{\rm cm}$ and $3.97 \times 10^{14}\,{\rm cm}$, respectively.  
At later times, the signal is increasingly sourced by positrons stopping at larger radii.
Because the CSM density decreases outward, the available column density $\propto \int dr \rho(r)$ becomes smaller, so that an increasing fraction of the injected positrons escapes before slowing
to sufficiently low energies to annihilate locally and contribute to
the narrow 511~keV line.
The line flux therefore falls rapidly beyond the peak.
Accordingly, the 511~keV signals generally peak at later times and last longer with 
denser and more extended CSM profiles; see Fig.~\ref{fig:flux-app} in Supplemental Material.

Importantly, the 511~keV signal is expected to peak before other electromagnetic emissions caused by the interaction of supernova shock with CSM.
This is because DPs and decay positrons propagate relativistically, much faster than the typical supernova shock velocity of $\sim 0.03~c$.  
Taking SN~2023ixf as an example, the earliest observed emissions are recorded only at $\sim 1$ day after the core-collapse~\cite{Zimmerman:2023mls,Cheng:2026ugs,Itagaki,amateur_TNS,Li:2023vux}.
The positrons responsible for the peak signal therefore reach and annihilate in the relevant CSM region well before the shock arrives.
While at sufficiently late times, the  
tail of the signal may overlap with emissions caused by the supernova shock, we only use flux around the peak time to derive the projected sensitivity presented below and neglect any potential emissions of sub-MeV $\gamma$-rays due to shock-CSM interaction~\cite{Sarmah:2022vra,Murase:2023chr,Sarmah:2023xrm,Kimura:2024lvt}. 

\begin{figure}[t!]
\begin{centering}
\includegraphics[width=1.0\columnwidth]{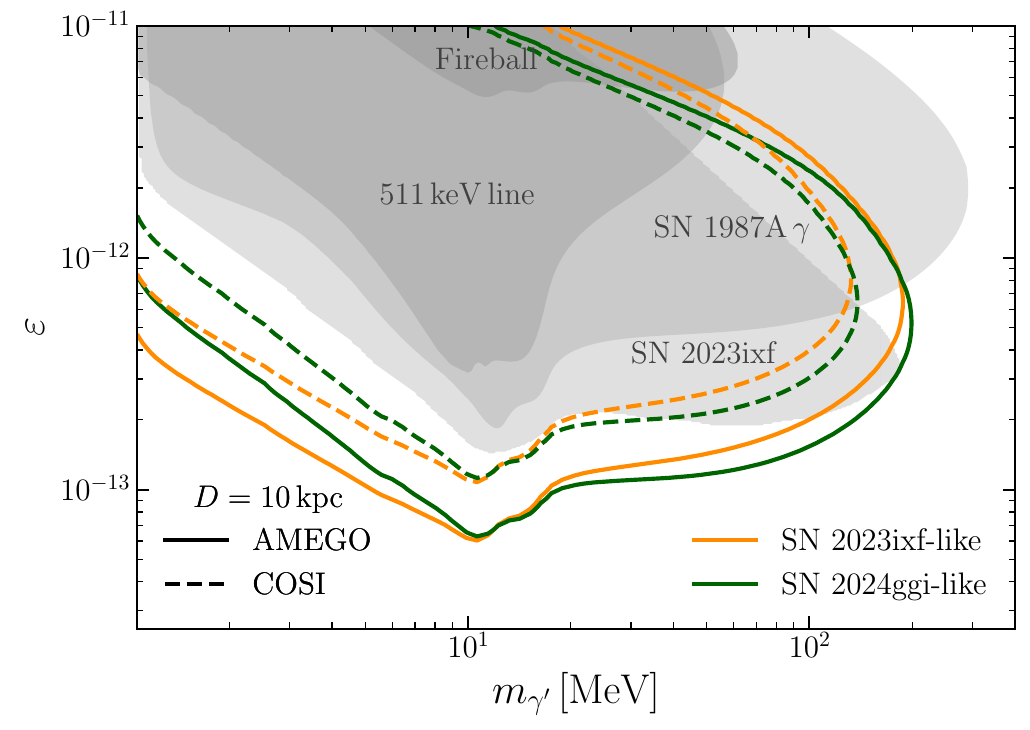}
\end{centering}
\caption{
Projected 511 keV sensitivity to the DP parameter space for a CCSN at distance $D = 10$\,kpc. 
The solid and dashed lines correspond to AMEGO and COSI, respectively, with the benchmark profiles motivated by SN~2023ixf (orange) and SN~2024ggi (green).}
\label{fig:sensitivity_realistic}
\end{figure}

\textit{\textbf{Projected Sensitivity.}---}
For each parameter point $(m_{\gamma^\prime},\varepsilon)$, we calculate the time-dependent 511~keV line flux using Eq.~\eqref{eq:flux}. 
To derive the projected sensitivity, we define the signal window, $T_{\rm win}$, as the time interval over which the 511\,keV flux exceeds $5\%$ of its maximum value.
The flux is then averaged over this window
and compared with the rescaled $3\sigma$ narrow-line sensitivity
\begin{equation}
 S(T_{\rm win})
 =
 S_{\rm ref}
 \left(\frac{T_{\rm ref}}{T_{\rm win}}\right)^{1/2}\,,
\end{equation}
where $S_{\rm ref}$ is the $3\sigma$ narrow line sensitivity for a reference exposure time $T_{\rm ref}$.
For COSI, we adopt a $3\sigma$ narrow-line sensitivity at 511~keV of approximately $1.2\times 10^{-5}\,{\rm ph\,cm^{-2}\,s^{-1}}$ for a two-year exposure~\cite{Tomsick:2019wvo,Tomsick:2023aue}.
For AMEGO, we take a deeper $3\sigma$ narrow-line sensitivity of approximately $9\times 10^{-6}\,{\rm ph\,cm^{-2}\,s^{-1}}$ for an exposure of
$10^6\,{\rm s}$~\cite{Caputo:2017sjw,Caputo:2022xpx}.
Taking $D=10$~kpc, the resulting projected sensitivities to the kinetic mixing $\varepsilon$ as a function of the DP mass $m_{\gamma^\prime}$ are shown in
Fig.~\ref{fig:sensitivity_realistic} for the benchmark profiles motivated by SN~2023ixf (orange) and SN~2024ggi (green), with COSI (dashed) and AMEGO (solid).   
Clearly, a next Galactic supernova surrounded by dense CSM can probe parameter space beyond leading constraints. 
Results using other reference CSM configurations are given in the Supplemental Material.

Ref.~\cite{Cheng:2026ugs} describes that, in part of the parameter space
probed by our projected sensitivity, energy deposition from the injected
electrons and positrons can heat the CSM above $\sim 3000\,{\rm K}$ and
increase its ionization fraction.
The enhanced free-electron abundance promotes Coulomb thermalization,
thereby suppressing positronium formation and increasing the fraction of
thermalized positrons~\cite{Guessoum:2005cb}.
These thermalized positrons subsequently annihilate with free or bound electrons, contributing directly to the 511~keV line (middle channel in Fig.~\ref{fig:AnnihilationChannels}).
Since Eq.~\eqref{eq:q511emissivity} includes only the $f_{\rm pPs}=1/4$ para-positronium fraction in the line signal, shifting annihilations from positronium formation toward direct two-photon annihilation can only increase the predicted 511~keV flux.
Our projected sensitivity based on the cold-CSM treatment is therefore conservative with respect to this heating effect.

\textit{\textbf{Summary and discussion.}---}
We have shown that dense CSM around CCSN progenitors provides a new target for transient 511~keV searches for visibly decaying FIPs. 
For the DP benchmark, positrons injected by DP decays either annihilate in flight or slow down in the CSM and subsequently form positronium or annihilate directly. 
The resulting photons can escape when produced outside the 511~keV photosphere and arrive as a geometrically broadened signal lasting $\sim10^3$--$10^5\,{\rm s}$. 
For a Galactic CCSN at $D=10\,{\rm kpc}$ with CSM profiles motivated by SN~2023ixf and SN~2024ggi, the projected COSI and AMEGO sensitivities extend beyond existing SN bounds in the MeV-scale DP parameter space.

This 511~keV probe is complementary to the CSM dust-sublimation signature proposed in Ref.~\cite{Cheng:2026ugs}, which can produce a rapid spectral transition of the progenitor emission from infrared-excess to optical/UV, based on the criterion that the CSM temperature gets heated over $3000\,{\rm K}$ at typical locations where the CSM dust resides.  
For much of the parameter space within our projected 511~keV reach, this criterion is also satisfied, so a nearby Galactic CCSN could exhibit both a dust-destruction signature and a transient annihilation line. 
As these complementary signals trace different consequences of the same DP decay, coincident observations can provide a
distinctive correlated evidence of FIP visible decay.

Dense CSM can also modify prompt $\gamma$-ray and fireball probes of visible FIP decays~\cite{Kazanas:2014mca,DeRocco:2019njg,Diamond:2023scc}. 
For prompt photons from final-state radiation, $\gamma'\to e^-e^+\gamma$, the production spectrum is governed by the decay kinematics, but photons produced inside the CSM can be 
reprocessed before escape. 
Likewise, energy transfer to the CSM can suppress the formation and expansion of 
a fireball, for which negligible baryon loading is usually assumed. 
With the presence of dense and cold CSM, the deposited energy is insufficient to accelerate the baryonic materials to relativistic velocities or heat the gas to the pair production scale to produce a fireball. 
Existing bounds and future projections that neglect the CSM may therefore require a dedicated treatment of the decay location, photon transport, and energy exchange with the gas.

Finally, dense CSM can also modify the Galactic 511~keV bound from positron injection~\cite{DeRocco:2019njg,Calore:2021lih}, 
which assumes that most positrons produced by FIP decays outside the progenitor environment enter the interstellar medium. 
For CCSNe surrounded by dense CSM, however, a fraction of these positrons can instead slow down and annihilate locally, reducing their contribution to the Galactic positron population. 
The Galactic 511 keV bound may therefore be weaker than conventionally estimated.
A quantitative reassessment would require incorporating the occurrence rate and distribution of CSM properties among Galactic CCSNe, which we leave for future work.

\bigskip
\textit{\textbf{Acknowledgments.---}} 
We are grateful to Thomas Janka for granting us the use of their SN models. 
YC is supported by the National Research Foundation of Korea (NRF) Grant RS-2023-00211732, 
by the Samsung Science and Technology Foundation under Project Number SSTF-BA2302-05, by the POSCO Science Fellowship of POSCO TJ Park Foundation, and by the NRF Grant RS-2024-00405629. 
CFK and SY are supported by IBS under the project code, IBS-R018-D1.
SY is supported by Basic Science Research Program through NRF funded by the Ministry of Education (No. RS-2026-25571203).
YHL and MRW acknowledge support of the National Science and Technology Council, Taiwan under Grant Nos.~115-2112-M-001-037-MY3, 111-2628-M-001-003-MY4, No.~115-2112-M-001-054-MY5, and Academia Sinica under Project No.~AS-IV-114-M04. 
YHL further acknowledges support from the Shui-Chin Lee Foundation through the Shui-Chin Lee Fellowship.
MRW also acknowledges support of the Physics Division of the National Center for Theoretical Sciences, Taiwan.

\bibliographystyle{apsrev4-1}
\bibliography{ref}

\clearpage

\appendix
\onecolumngrid
\section*{End Matter}
\twocolumngrid

\section{Positron stopping in the CSM}\label{app:stopping}

To determine where a positron stops in the CSM, 
we describe the energy loss through propagation using the continuous-slowing-down approximation (CSDA) following Ref.~\cite{Cheng:2026ugs}, 
in which the positron energy decreases continuously along its trajectory.
The mass stopping power is defined as
\bea
S(E)=-\frac{1}{\rho}\frac{\mathrm{d}E}{\mathrm{d}x} \,,
\eea
such that $\rho S(E)$ gives the energy loss per unit distance.

For a fixed chemical composition, the mass stopping power is independent of the local density since $dE/dx\propto \rho$.
The stopping condition can therefore be expressed naturally in terms of the traversed column density.
For a positron injected at radius $r_{\rm d}$ with initial energy $E_i$, the stopping radius $r_{\rm s}(r_{\rm d},E_i)$ is determined by solving 
\bea
\int_{r_{\rm d}}^{r_{\rm s}} dr\,\rho(r) = \int_{E_{\rm stop}}^{E_i} \frac{dE}{S(E)}\,, 
\label{eq:stopping-column}
\eea
where
$E_{\rm stop}=m_e+K_{\rm stop}$ with $K_{\rm stop}=10\,{\rm keV}$. 
If the solution to Eq.~\eqref{eq:stopping-column} exists, the positron stops within the CSM at $r_{\rm s}(r_{\rm d},E_i)$, and the corresponding stopping distance is $\delta r_{\rm stop}=r_{\rm s}-r_{\rm d}$. 
Otherwise, the positron escapes from the CSM and subsequently free-streams. 
Below $E_{\rm stop}$, 
positrons continue to lose energy due to other low-energy atomic processes, which are treated explicitly as discussed in the Section ``Positronium formation and decay into 511 keV lines'' in this End Matter.

The stopping length calculated above also allows to compute the stopping time
\bea
t_{\rm stop} = \int_{r_{\rm d}}^{r_{\rm s}} \frac{dr}{\beta_{e^+}(r)} \sim \delta r_{\rm stop}\,,
\eea
where $\beta_{e^+}(r)$ is the positron velocity.
The final expression provides an order-of-magnitude estimate since most of the propagation distance is accumulated while the positron remains relativistic.

\section{Annihilation in flight and positron survival}\label{app:survival}

During the slowing-down stage, a positron may annihilate in flight with an ambient electron and produce a broad continuum $\gamma$-ray spectrum rather than a narrow 511~keV line~\cite{Beacom:2005qv,DelaTorreLuque:2024zsr}.
This continuum may provide a complementary signature for $\gamma$-ray telescopes.
In this work, however, we focus on the narrow 511~keV signal and leave a detailed study of the in-flight annihilation continuum to future work.
The annihilation cross section for a positron on an electron at rest is~\cite{Dirac:1930bga,Beacom:2005qv,DelaTorreLuque:2024zsr}
\begin{equation}
\begin{aligned}
 \sigma_{\rm ann}(E)
 &=
 \left(\frac{\alpha}{m_e}\right)^2
 \frac{\pi}{\gamma_e+1}
 \\
 &\quad\times
 \left[
 \frac{\gamma_e^2+4\gamma_e+1}{\gamma_e^2-1}
 \ln\left(\gamma_e+\sqrt{\gamma_e^2-1}\right)
 \right.
 \\
 &\qquad\left.
 -\frac{\gamma_e+3}{\sqrt{\gamma_e^2-1}}
 \right],
\end{aligned}
\end{equation}
where $\gamma_e=E/m_e$ is the positron Lorentz factor.
The fraction of positrons annihilating over a distance $dx$ during the stopping process satisfies
\begin{equation}
 \frac{dN}{N}
 =
 -n_e\sigma_{\rm ann}\,dx
 =
 dE\,\frac{n_e \sigma_{\rm ann}}{\rho S(E)}.
\end{equation}

For a neutral H/He gas, let $X$ and $Y$ be the mass fractions of H and He, respectively.
With $n_{\rm H}=X\rho/m_p$ and $n_{\rm He}=Y\rho/4m_p$, the electron density per unit mass is
\begin{equation}
 \frac{n_e}{\rho}
 =
 \frac{1}{m_p}\left(X+\frac{Y}{2}\right).
\end{equation}
The survival probability is therefore
\begin{equation}
 P_{\rm surv}(E_i\to E_{\rm stop})=
 \exp\left[-\int_{E_{\rm stop}}^{E_i}dE\,
 \frac{\sigma_{\rm ann}(E)}
 {S(E)m_p}
 \left(X+\frac{Y}{2}\right)\right].
 \label{eq:app_survival}
\end{equation}
Both the in-flight annihilation rate and the continuous energy-loss rate scale linearly with the gas density, so their leading density dependence cancels in $P_{\rm surv}$.
For a fixed composition, the survival probability therefore depends primarily on the initial positron energy.

\section{Positronium formation and decay into 511\,keV lines}

\begin{figure}[t!]
\begin{centering}
\includegraphics[width=1.0\columnwidth]{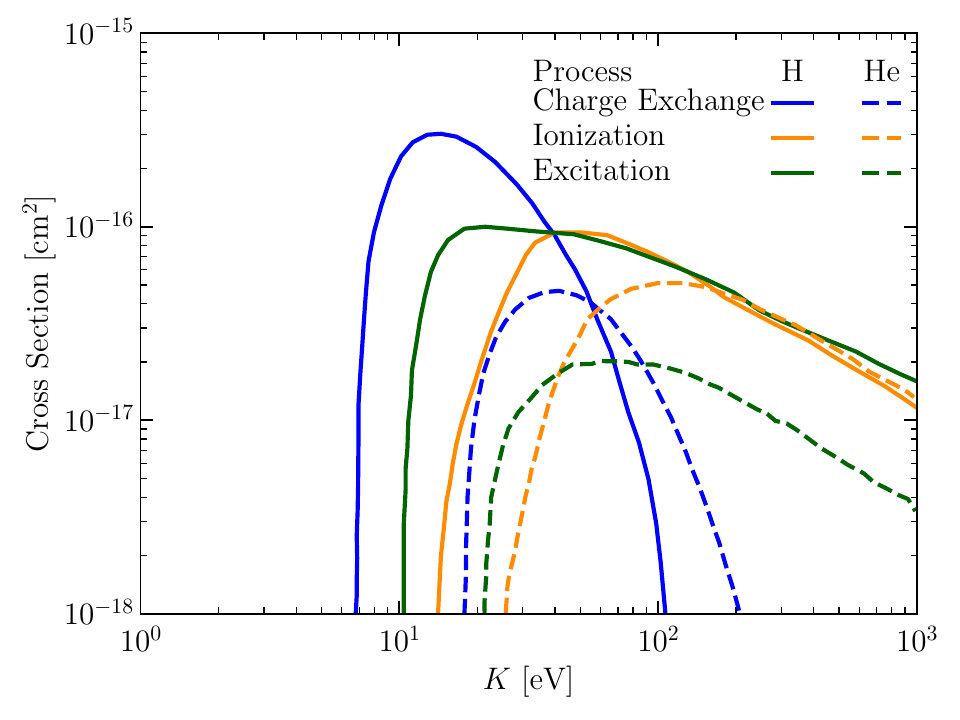}
\caption{
Cross sections used in the low-energy Monte Carlo
calculation as functions of the positron kinetic energy. Solid
and dashed curves correspond to hydrogen and helium, respectively; colors distinguish charge exchange (blue), ionization (orange), and excitation (green). Data are taken from
Ref~\cite{Guessoum:2005cb}.
}
\label{fig:CrossSection}
\end{centering}
\end{figure}

Positrons that survive in-flight annihilation enter the
eV–keV regime after losing most of their kinetic energies.
At this stage, positrons can interact with ambient electrons through several channels as illustrated in Fig.~\ref{fig:AnnihilationChannels}, 
including charge exchange with neutral atoms, radiative recombination with free electrons, direct annihilation with free electrons, and direct annihilation with bound electrons in neutral atoms~\cite{Guessoum:2005cb,Prantzos:2010wi}.

Charge exchange and radiative recombination first form positronium, which subsequently decays on a timescale negligible for the present analysis. 
By $C$-parity conservation, the singlet state, para-positronium, constitutes $1/4$ of the formed positronium and decays predominantly into two 511\,keV photons.
The triplet state, ortho-positronium, constitutes the remaining $3/4$ and decays into three photons with a continuum spectrum below $511\,{\rm keV}$.

Direct annihilation with either free or bound electrons produces two 511\,keV photons without forming an intermediate positronium state.
The relative importance of these channels relies on their respective cross sections, the ionization fraction and the chemical composition of CSM gas, the gas temperature, and the positron kinetic energy during the final stage of slowing. 

To determine the fate of positrons during this final stage, we employ a Monte Carlo simulation including discrete charge-exchange (CE), excitation, and ionization events, using the cross sections shown in Fig.~\ref{fig:CrossSection} and following Refs.~\cite{Guessoum:2005cb,1979ApJ...228..928B}.
For target species $s\in\{{\rm H},{\rm He}\}$ and process $a\in\{{\rm CE},{\rm exc},{\rm ion}\}$, the probability of the next event at kinetic energy $K$ is given by
\begin{equation}
 P_{a,s}(K)=
 \frac{n_s\sigma_{a,s}(K)}
 {\sum_{s',a'}n_{s'}\sigma_{a',s'}(K)}.
 \label{eq:mc_probability}
\end{equation}
Effectively, the positronium formation probability obeys the following recursion relation
\begin{equation}
\begin{aligned}
 F_{\rm Ps}(K)&=\sum_s P_{{\rm CE},s}(K) \\
 &\quad+\sum_s P_{{\rm exc},s}(K)
 F_{\rm Ps}(K-\Delta K_{{\rm exc},s}) \\
 &\quad+\sum_s P_{{\rm ion},s}(K)
 F_{\rm Ps}(K-\Delta K_{{\rm ion},s}),
\end{aligned}
 \label{eq:fps_recursion}
\end{equation}
with $F_{\rm Ps}=0$ once no charge-exchange channel remains open.

Instead of solving directly
Eq.~\eqref{eq:fps_recursion}, we inject $N_{e^+}$ positrons into our Monte Carlo simulation with initial kinetic energy $K=K_i=10$~keV.
Charge exchange terminates a trajectory by forming positronium and is allowed when $K$ is above $6.8\,$eV for H and $17.8\,$eV for He. 
Otherwise, excitation reduces $K$ by $10.2\,$eV (H) or $21.2\,$eV (He), while ionization reduces it by $I_s+\varepsilon_{\rm ej}$, with $I_{\rm H}=13.6\,$eV, $I_{\rm He}=24.6\,$eV, and $\varepsilon_{\rm ej}\simeq I_s/4$.
The process is repeated until charge exchange occurs or all CE channels close.

For $N_{e^+}$ simulated trajectories, the positronium-formation fraction is estimated as 
\begin{equation}
 f_{\rm Ps}(K_i)
 =
 \frac{N_{\rm CE}}{N_{e^+}},
\end{equation}
where $N_{\rm CE}$ counts trajectories ending in charge exchange.
For neutral CSM with H/He mass fractions $X=0.7$ and $Y=0.3$, we obtain $f_{\rm Ps}\simeq0.97$, nearly independent of the initial positron energy and the gas density.

The time scale required for positronium formation can be estimated as follows. 
For a positron with kinetic energy $K_+\sim \mathcal{O}(10)\,{\rm eV}$, it has a charge-exchange cross section of order $\sigma_{\rm CE}\sim 10^{-16}\,{\rm cm^2}$.
For a representative dense-CSM number density $n_{\rm tot}\sim 10^{11}\,{\rm cm^{-3}}$, the formation timescale reads $t_{\rm form} \sim 1/{n_{\rm tot} \sigma_{\rm CE} v_{e^+}} \approx 10^{-4}\,$s. 
This timescale is much shorter than both the 
slowing-down time and the macroscopic timescale of the 511~keV line signal.

For positrons that survive from forming positronium according to the above calculations, they continue losing energy toward thermal equilibrium with the ambient gas. 
For the cold and mostly neutral CSM considered here
with a typical temperature $T\lesssim 3000\,{\rm K}$~\cite{Dessart:2017pfi,Jacobson-Galan:2021pki}, 
because the thermal positron energies lie well below the charge-exchange threshold, charge exchange after thermalization would be strongly suppressed. 
As a result, these positrons eventually annihilate into 511\,keV photons through direct interactions with bound electrons in neutral atoms. 
Although this corresponding cross section is approximately one to four orders of magnitude smaller than the charge-exchange cross section~\cite{Guessoum:2005cb,Prantzos:2010wi}, the high CSM density still ensures that this annihilation occurs on a short timescale.

\clearpage
\newpage
\maketitle
\onecolumngrid

\setcounter{equation}{0}
\setcounter{figure}{0}
\setcounter{section}{0}
\setcounter{table}{0}
\setcounter{page}{1}
\makeatletter

\renewcommand{\theequation}{S\arabic{equation}}
\renewcommand{\thefigure}{S\arabic{figure}}
\renewcommand{\thetable}{S\arabic{table}} 

\begin{center}
\textbf{\large 
Probing Feebly Interacting Particles with 511 keV Line 
from Circumstellar Medium of Supernovae}

\vspace{0.05in}
{ \it \large Supplemental Material}\\ 
\vspace{0.05in}
{Yu Cheng, Chui-Fan Kong, Yen-Hsun Lin, Meng-Ru Wu, and Seokhoon Yun}
\end{center}

\twocolumngrid

\section{Constraints from  
general CSM density profiles}
\label{sec:rho-profile}

\begin{figure}[t!]
\begin{centering}
\includegraphics[width=.95\columnwidth]{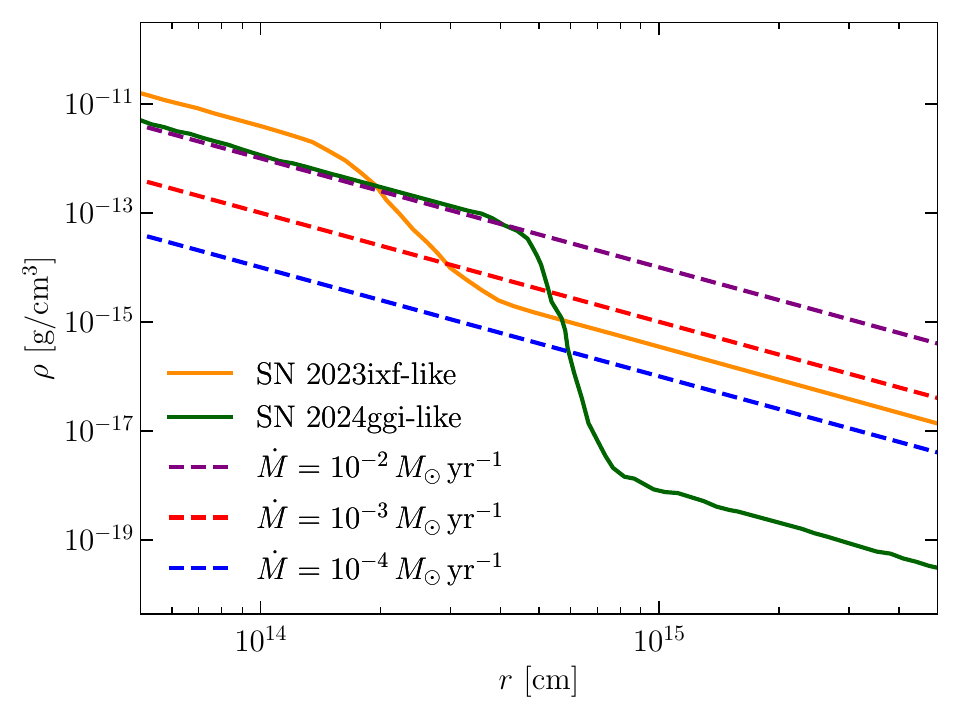}
\end{centering}
\caption{
CSM density profiles adopted in this work, the observationally motivated SN 2023ixf-like (orange) and SN 2024ggi-like (green) profiles, and a steady RSG wind with $\dot{M}=10^{-2}\, M_{\odot}\,\mathrm{yr}^{-1}$ (purple dashed), $10^{-3}\,M_{\odot}\, \mathrm{yr}^{-1}$ (red dashed), and $10^{-4}\,M_{\odot}\,\mathrm{yr}^{-1}$ (blue dashed), assuming $v_w=50 \mathrm{~km} \mathrm{~s}^{-1}$ (purple).
}
\label{fig:density}
\end{figure}

\begin{figure}[t!]
\begin{centering}
\includegraphics[width=.95\columnwidth]{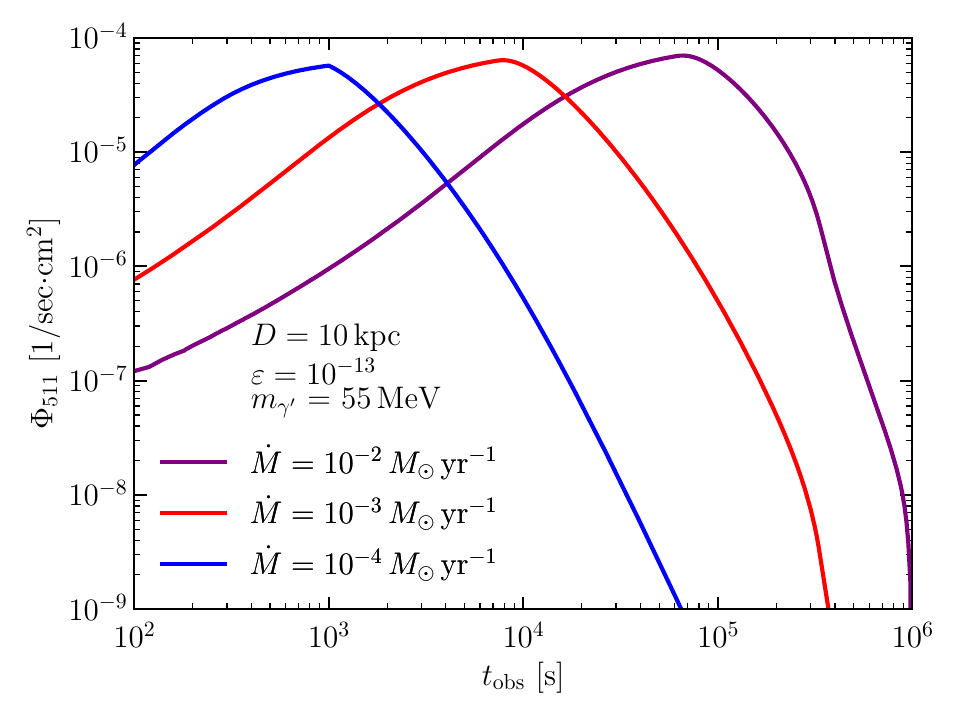}
\end{centering}
\caption{
Time-dependent 511 keV flux for a Galactic CCSN at 
$D=10 \, {\rm kpc}$, with $m_{\gamma^{\prime}}=55 \,{\rm MeV}$ and $\varepsilon=10^{-13}$. Three mass loss rates of $\dot M=10^{-2},\,10^{-3}$, and $10^{-4}\,M_\odot\,\mathrm{yr}^{-1}$ are shown in purple, red, and blue curves, respectively.
}
\label{fig:flux-app}
\end{figure}

\begin{figure}[t!]
\begin{centering}
\includegraphics[width=.95\columnwidth]{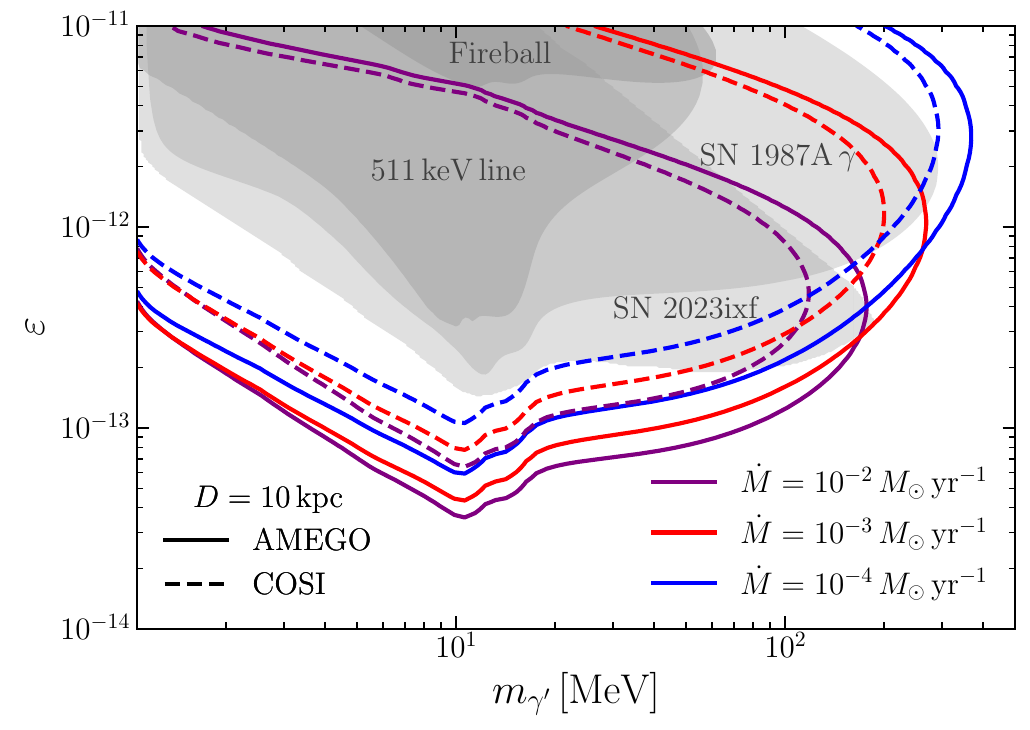}
\end{centering}
\caption{\label{fig:sensitivity}
Projected 511 keV sensitivity to the DP kinetic mixing for a CCSN at $D = 10$\,kpc. 
The purple, red, and blue curves correspond to mass loss rates of $\dot M=10^{-2},\,10^{-3}$ and $10^{-4}\,M_\odot\,\mathrm{yr}^{-1}$, respectively.
Solid curves show the projected AMEGO sensitivity,
while dashed curves show the corresponding COSI sensitivity.
}
\label{fig:constraint-appendix}
\end{figure}

In the main text, we adopt the SN 2023ixf-like and SN 2024ggi-like CSM density profiles, shown as the solid curves in Fig.~\ref{fig:density}. 
Both profiles feature a compact and high density inner CSM extending to radii of order $10^{14}\,{\rm cm}$.
The SN 2023ixf-like profile is more centrally concentrated, while the SN 2024ggi-like profile remains comparatively dense out to $r\sim 4\times10^{14}\,{\rm cm}$ before declining sharply into its low density outer component, resulting in a more sharply truncated CSM. 

For comparison, we consider in this Supplemental Material additional CSM profiles described by a generic steady wind model, whose density is given by 
\begin{equation}
    \rho 
    =
    \frac{\dot M}{4\pi r^2 v_{\rm w}},
\end{equation}
where $\dot M$ is the mass loss rate and $v_{\rm w}$ is
the wind velocity.
For illustration, we take
three general CSM mass loss rates, namely,
$\dot M = 10^{-2}\,M_\odot\,{\rm yr}^{-1}, 10^{-3}\,M_\odot\,{\rm yr}^{-1}$, and $10^{-4}\,M_\odot\,{\rm yr}^{-1}$.
The three density profiles are shown by dashed curves in Fig.~\ref{fig:density}.
The corresponding 511~keV line fluxes and the projected sensitivities with COSI and AMEGO
are shown in Figs.~\ref{fig:flux-app} and ~\ref{fig:constraint-appendix}, respectively.
One can also find that projected sensitivity is stronger than the current best bound over a broad range of mass loss rates, $\dot{M}/(\, M_\odot\,{\rm yr}^{-1})\in [10^{-4},10^{-2}]$. 
A higher $\dot{M}$ profile yields a more stringent constraint because the denser CSM stops a larger fraction of the positrons before they escape to infinity, thereby enhancing the annihilation signal.
For CSM models with $\dot{M} < 10^{-4}\, M_\odot\,{\rm yr}^{-1}$\, 
the photospheric radius lies below the typical radius of red supergiant progenitor stars.
We therefore do not consider these cases in this work.

\end{document}